# Ultrafast Charge-Doping via Photo-Thermionic Injection in van der Waals Devices

Yiliu Li[1†], Esteban Rojas-Gatjens[1†], Yinjie Guo[2], Birui Yang[2], Dihao Sun[2], Luke Holtzman[3], Juseung Oh[1], Katayun Barmak[3], Cory R. Dean[2], James C. Hone[4], Nathaniel Gabor[5], Eric A. Arsenault[1*], Xiaoyang Zhu[1*]

[1]Department of Chemistry, Columbia University, New York, NY 10027, USA
[2]Department of Physics, Columbia University, New York, NY 10027, USA
[3]Department of Applied Physics and Applied Mathematics, Columbia University, New York, NY 10027, USA
[4]Department of Mechanical Engineering, Columbia University, New York, NY 10027, USA
[5]Department of Physics and Astronomy, University of California, Riverside, CA 92521, USA

## Abstract

Van der Waals (vdW) heterostructures of two-dimensional (2D) materials have become a rich playground for the exploration of correlated quantum phases, and recent studies have begun to probe their non-equilibrium dynamics under femtosecond laser excitation. In a time-resolved experiment, optical excitation of the multilayer structure can lead not only to rich dynamic responses from the target layers, such as moiré interfaces, but also to additional device functionality from the layer degree of freedom. Here, we investigate optical excitation in a prototypical moiré device of dual-gated twisted $WSe_2$ bilayers, with few-layer graphite gates and hexagonal boron nitride spacers. We establish an ultrafast photodoping mechanism in the moiré bilayer from photo-thermionic emission of the graphite gates. Using transient reflectance experiments, we reveal photo-induced hole injection evidenced by: (i) a shift of gate voltages at which optical signatures of correlated insulators are observed, (ii) a persistent optical signature indicative of charge diffusion at microsecond timescales and local charge buildup from pulse-to-pulse accumulation, and (iii) photoinduced absorption due likely to transient formation of correlated insulators. We further demonstrate that the injected holes can be selectively controlled by tuning the excitation energy, fluence, and gate bias.

† These authors contributed equally to this work.
* Corresponding author. Email: eaa2181@columbia.edu, xyzhu@columbia.edu

## Introduction

A plethora of quantum phenomena have been reported in vdW moiré structures, including Mott insulators and generalized Wigner crystals [1–4], integer and fractional quantum anomalous Hall states [5–7], and superconductivity [8,9]. These discoveries have motivated experiments to probe their non-equilibrium responses that can reveal the underlying physics [10–13]. A typical time-resolved experiment uses an ultrashort pulse to photo-excite quasi-particles across the correlated insulator gap or pseudo gap, followed by a probe pulse resonant with an excitonic transition to monitor pump-induced changes in the dielectric environment. In our recent work [10–12], we employed transient reflectance measurements to study the disordering and reordering dynamics of a range of correlated phases in transition metal dichalcogenide (TMD) moiré bilayers, including angle-aligned $WSe_2/WS_2$ [10,11] and twisted bilayer $MoTe_2$ [12]. To correctly interpret such measurements and, when possible, exploit the multifunctionality of the vdW structure, it is essential to resolve the full photo-response of the device beyond the target moiré interfaces. For example, the layer degree of freedom under photo-excitation may be used to control the generation and propagation of phonon wavepackets [14,15], the dynamic screening of excitons [16], and the separation of electronic and phononic melting of quantum phases [17]. Additionally, we noted the possibility of photo-thermionic hole injection from graphite gates at sufficiently high excitation densities in the transient reflection spectra from a dual-gated $WSe_2/WS_2$ device [10].

In a vdW device, carrier doping is typically achieved with a dual-gated structure in which the active moiré layers are encapsulated by top and bottom hexagonal boron nitride (hBN) and fl-Gr gate electrodes. While optical excitation targets the moiré layers, it also excites the semi-metallic gate electrodes to create a non-equilibrium carrier distribution, which rapidly thermalizes through electron-electron scattering, forming a hot-carrier distribution [18]. At high optical excitation density, hot carriers in fl-Gr can possess sufficient energies to overcome the energy barrier of the hBN dielectric, leading to interlayer carrier transfer, a process known as photo-thermionic injection [19,20]. This mechanism underlies many photovoltaic and photodetector devices based on similar heterostructures [21–23], where graphene is typically paired with a semiconductor (e.g., TMD or $TiO_2$) that acts as a hot-carrier acceptor. The hBN dielectric could also act as a hole acceptor either through tunneling or photo-thermionic injection mechanisms, depending on the photon energy and voltage bias applied [24]. The parallels between vdW stacks in optoelectronic applications and vdW heterostructures used to realize quantum phases suggest that photo-thermionic hole injection from the graphite gates could transiently modulate the doping density of a moiré superlattice.

In this work, we demonstrate the photodoping of a dual-gated twisted $WSe_2$ bilayer device (D1), with twist angle $\theta = 58.1\pm0.1^{o}$. Reproducible results from a second $WSe_2$ bilayer device (D2), with twist angle $\theta = 58.2\pm0.1^{o}$ are shown in Supporting Information [25], Fig. S1, S2. A steady-state electron population is introduced by gate doping, while holes are injected on picosecond timescales through the photo-thermionic emission process. The injected holes recombine with the electrostatically doped electrons on ultrafast timescales, thus transiently shifting the doping density of the system. We resolve these doping density alterations over time via transient reflectance spectroscopy, where changes in the dielectric environment are monitored via exciton sensing [10–12]. Specifically, we observe three distinct spectral signatures that evidence photo-induced hole injection: (i) the photo-bleaching (PB) signature for the Mott insulator state shifting towards positive gate voltages, (ii) a persistent ($\geq 2.5$ μs) feature signaling a charge diffusion at microsecond timescales and local charge build up from pulse-to-pulse accumulation, and (iii) the emergence of a photo-induced absorption (PA) feature suggesting a photo-induced metal-to-insulator transition. Additionally, we control the transient charge carrier doping by tuning the laser wavelength, excitation fluence, and gate voltage. Here, we establish the transient hole doping mechanism from the fl-Gr electrodes to $WSe_2/WSe_2$ homobilayers and present a quantitative and systematic characterization of photo-thermionic charge injection, while a companion manuscript focuses on photo-induced metal-to-insulator transition in the $WS_2/WSe_2$ system [26]. All measurements are carried out at 4.6K. Increasing temperature is expected to affect multiple processes, including photo-thermionic injection efficiency, lateral charge diffusion coefficients, and charge recombination rates. We focus on the low-temperature regime where correlated electronic phases are most stable and spectroscopically obvious.

**Characterization of insulating phases under static and time-resolved measurements**

A schematic of the device structure is shown in Fig. 1a. The static electron or hole doping density is controlled by an applied bias between the TMD bilayer and the fl-Gr gates. Unless otherwise noted, the sample is symmetrically gated (same gate voltage $V_g$ for both top and bottom gates), where a positive (negative) $V_g$ corresponds to the electron (hole) doping. We monitor the oscillator strength of the lowest energy $WSe_2$ moiré exciton [27] as a function of $V_g$ and, thus, filling factor ($\nu_m$) of the moiré superlattice. The exciton oscillator strength ($f$) sensitively detects changes in the local dielectric environment ($\varepsilon_{eff}$), $f \propto 1/{\varepsilon_{eff}}^{\alpha}$, where the exponent $\alpha\,(\geq 1)$ is related to the form of Coulomb potential [28,29]. Specifically, upon formation of an insulating state, there is a significant reduction in $\varepsilon_{eff}$, which leads to an enhancement in $f$. In a transient reflectance measurement, a pump pulse with photon energy above the correlation gap or pseudo gap leads to disruption of correlation and gap closing [10,11,17], as reflected in an increase in $\varepsilon_{eff}$ and the corresponding decrease in $f$ detected by the exciton resonances in reflectance spectra by the probe pulse. Specifically, the transient reflectance is represented by $\Delta R/R_0$, where $\Delta R = R(\Delta t) - R_0$; $R(\Delta t)$ and $R_0$ are the reflectance at $\Delta t$ and the reflectance without pump excitation, respectively. The pump-induced reduction in the amplitude of the derivative-shaped reflectance spectrum, i.e., photobleach (PB), shows a characteristic flip in sign from that of $R_0$ as the probe photon energy increases across the resonance, illustrated at the bottom of Fig. 1a.

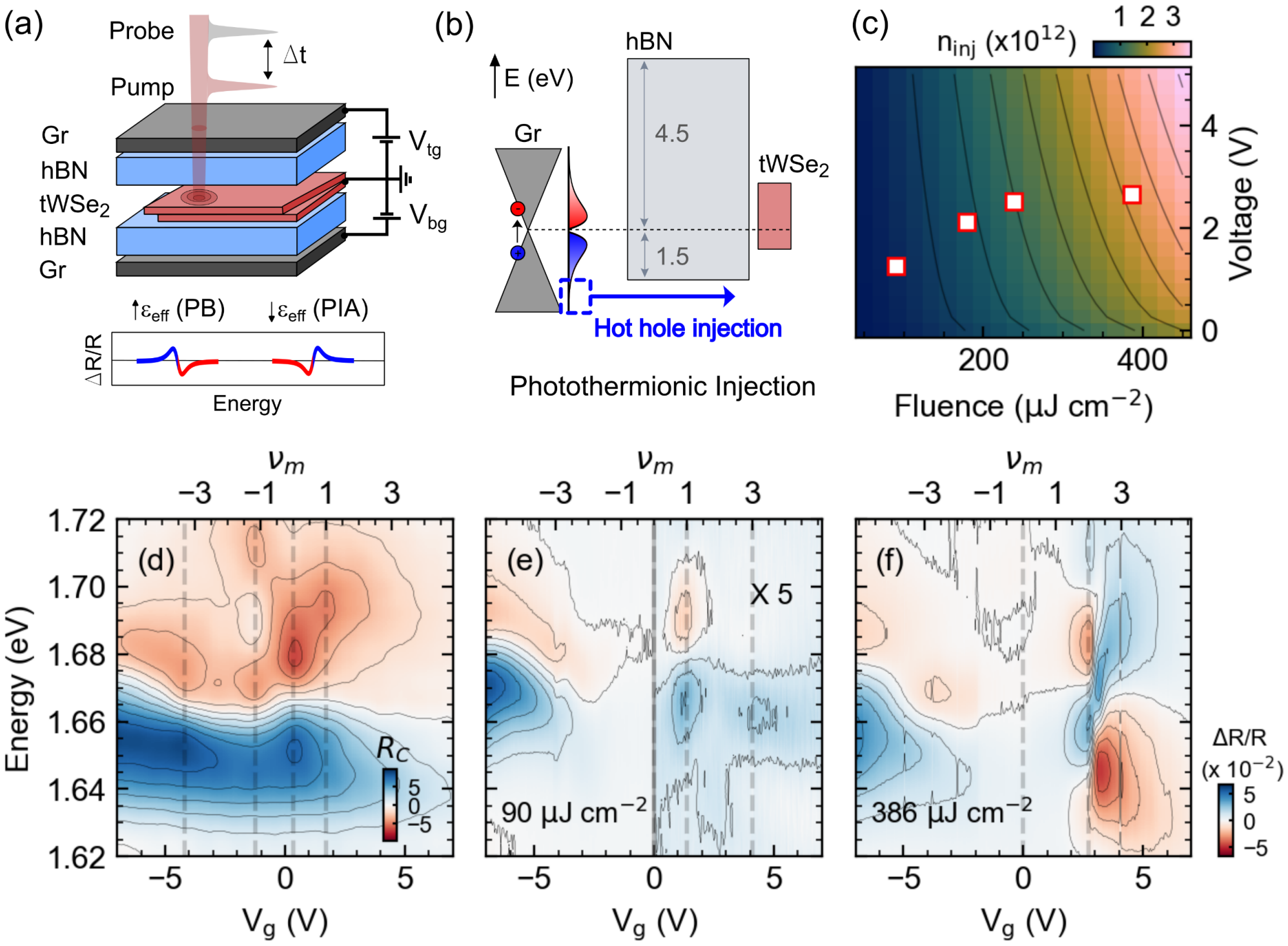

**Figure 1. Device architecture and characterization of insulating phases.** (a) Schematics of the device consisting of fl-Gr (Gr) top ($V_{tg}$) and bottom ($V_{bg}$) gates, hexagonal boron nitride (hBN) dielectrics, and the twisted $WSe_2$ bilayer with a twist angle of 58.1°. The cartoon at the bottom illustrates the expected photobleach lineshape in the transient reflectance experiment. (b) Schematic representation of hot hole injection via photo-thermionic emission. (c) Carrier density above the photo-thermionic injection barrier as a function of incident laser fluence and gate voltage, with the red squares corresponding to the experimental conditions in Fig. 2 (see the Supplementary Information for details). (d) Steady-state reflectance spectrum at T = 4.6 K of the twisted $WSe_2$ dual-gated device as a function of gate voltage ($V_g = V_{tg} = V_{bg}$). $R_C$ corresponds to the reflectance contrast defined as $(R_S - R_{BG})/R_{BG}$, where $R_S$ and $R_{BG}$ are the sample and background reflectance, respectively. Here, we use a highly electron-doped spectrum as the background to enhance the signal difference between the insulating and metallic states, see Methods for further information. Transient reflectance as a function of gate voltage (as labeled for each row) at a pump-probe delay of Δt = 10 ps at two pump fluences (e) 90 μJ $cm^{-2}$ and (f) 386 μJ $cm^{-2}$. Here, a pump photon energy of 0.954 eV was employed, and the probe energy was set to cover the lowest energy $tWSe_2$ moiré exciton. The contour lines divide the 2D map into ten equidistant segments between the maximum and minimum of the map.

The dynamic response of the vdW heterostructure is dominated by the melting and recovery of the correlated states in the moiré TMD bilayers at low excitation densities [10–12]. As we show below, at sufficiently high pump fluence, excitation in the fl-Gr electrodes can couple to the moiré TMD bilayers. Specifically, highly energetic carriers in the tail of the hot hole distribution can overcome the hBN barrier and transiently dope the TMD moiré bilayers. The energy offset between the Dirac point of graphene and the hBN valence band maximum (VBM) is approximately 1.3 eV [24,30], significantly smaller than the offset of 4.5 eV to the conduction band minimum (CBM), Fig. 1b. Therefore, photoexcitation of the Gr electrodes can inject holes, not electrons, into the TMD moiré bilayers. We carry out numerical simulations of hole injection as a function of gate bias and excitation fluence, as detailed in the Supporting Information [25]. The injection can come from both above-barrier thermionic emission and below-barrier tunneling, but the latter is negligible because of the thickness (40 nm) of the h-BN barrier. Fig. 1c shows the simulation results, with the red squares corresponding to the experimental conditions (gate bias $V_g$ and pulse fluence ρ).

We first show the differential reflectance spectra as a function of gate voltage, Fig. 1d, which reveals multiple states with integer $\nu_m$. These include the one electron filling ($\nu_m$ = 1) as well as the one and three hole filling ($\nu_m$ = -1, -3) states. In transient reflectance, the spectral features depend on the excitation fluence (ρ). We show the transient spectra as a function of gate voltage at a selected pump-probe delay of Δt = 10 ps at low (ρ = 90 μJ/cm$^2$, Fig. 1e) and high (ρ = 386 μJ/cm$^2$, Fig. 1f) excitation fluence. Since the excitation photon energy is below the bandgap of $WSe_2$, the pump pulse only excites across the correlation gap [31], inducing a disordering of the correlated insulator [10,11]. The corresponding decrease in the exciton oscillator strength appears as the characteristic PB feature in the transient reflectance spectra, as seen for the $\nu_m$ = 1 state at the low excitation fluence, Fig. 1e. At the high pump fluence, Fig. 1f, there is a pronounced shift in PB feature of the $\nu_m$ = 1 state towards more positive $V_g$. Moreover, there is a sign reversal of the reflectance spectra at higher $V_g$ (> 3 V), corresponding to a pump-induced decrease in $\varepsilon_{eff}$. The latter is consistent with the expected signature of a metal-to-insulator transition [30].

In the following, we use two pump photon energies (hν = 0.954 eV and 1.46 eV) to access different excitation conditions of the graphite gates. While thermionic injection is generated at either energy, we use the lower photon energy over a wider fluence range without saturation in density-dependent studies, whereas the higher photon energy is more efficient in generating hot carriers in the fl-Gr and used in repetition rate dependent measurements. We also take advantage of the distinct photonic effects at different photon energies to selectively excited the top or the bottom fl-Gr electrode.

**Transient doping of the moiré structure via photo-thermionic injection**

To establish the photo-thermionic emission mechanism for hole injection into the TMD bilayer, we carry out pump fluence-dependent measurements. Figs. 2a-h show transient reflectance spectra as a function of gate voltage at two selected waiting times, Δt = 10 ps (a-d) and -5 ps (e-h), and for pump fluences ρ = 90, 179, 238, and 386 μJ/cm$^2$, respectively. The pump probe delay time of -5 ps corresponds to a time separation of 2.5 μs between consecutive laser pulses, given the laser repetition rate of 400 kHz. The negative delay therefore probes residual signals persisting from the previous pump excitation pulse. Here, we focus on the electron doping side as the positive gate bias ($V_g$ > 0) assists hole injection from the photo-excited Gr electrodes. For comparison, the negative gate bias on the hole doping side inhibits hole injection from the Gr electrodes, and we find no evidence of excitation fluence-dependent changes to the hole doping levels, Fig. 1e,f.

At the low pump fluence of ρ = 90 μJ/cm$^2$, we observe the characteristic PB feature at $\nu_m$ = 1 at early delay times (Δt = 10 ps, Fig. 2a) and its disappearance at longer delay times (Δt = 2.5 μs, Fig. 2e), as expected from the melting and recovery of the correlated insulator [10–12]. At higher pump fluences, ρ = 179 - 386 μJ/cm$^2$, distinct new features are observed, signaling the emergence of a photo-injected hole population: (i) the PB feature associated with the insulating $\nu_m$ = 1 state shifts towards more positive gate voltages with increasing ρ (Fig. 2b-d); (ii) PB features indicative of increased $\varepsilon_{eff}$ persist for the longest delay time Δt = 2.5 μs as dictated by the repetition rate of the laser pulses, and (iii) a new spectral feature with a reversed sign appears, suggesting a decrease in $\varepsilon_{eff}$ or an increase in oscillator strength, i.e., photoinduced absorption (PIA). These features emerge from an interplay of the three characteristic timescales in the pump-probe measurement: the waiting time (Δt) between the pump and probe pulses (ps-ns), the inter-pulse time constant $\tau_{rep}$ (= $1/\nu_{rep}$ = 20-2.5 μs, where $\nu_{rep}$ = 50-400 kHz is the laser repetition rate), and pump on/off modulation time constant $\tau_{mod}$ (= $1/\nu_{mod}$ = 2 ms, where $\nu_{mod}$ = 0.5 kHz is the chopper frequency).

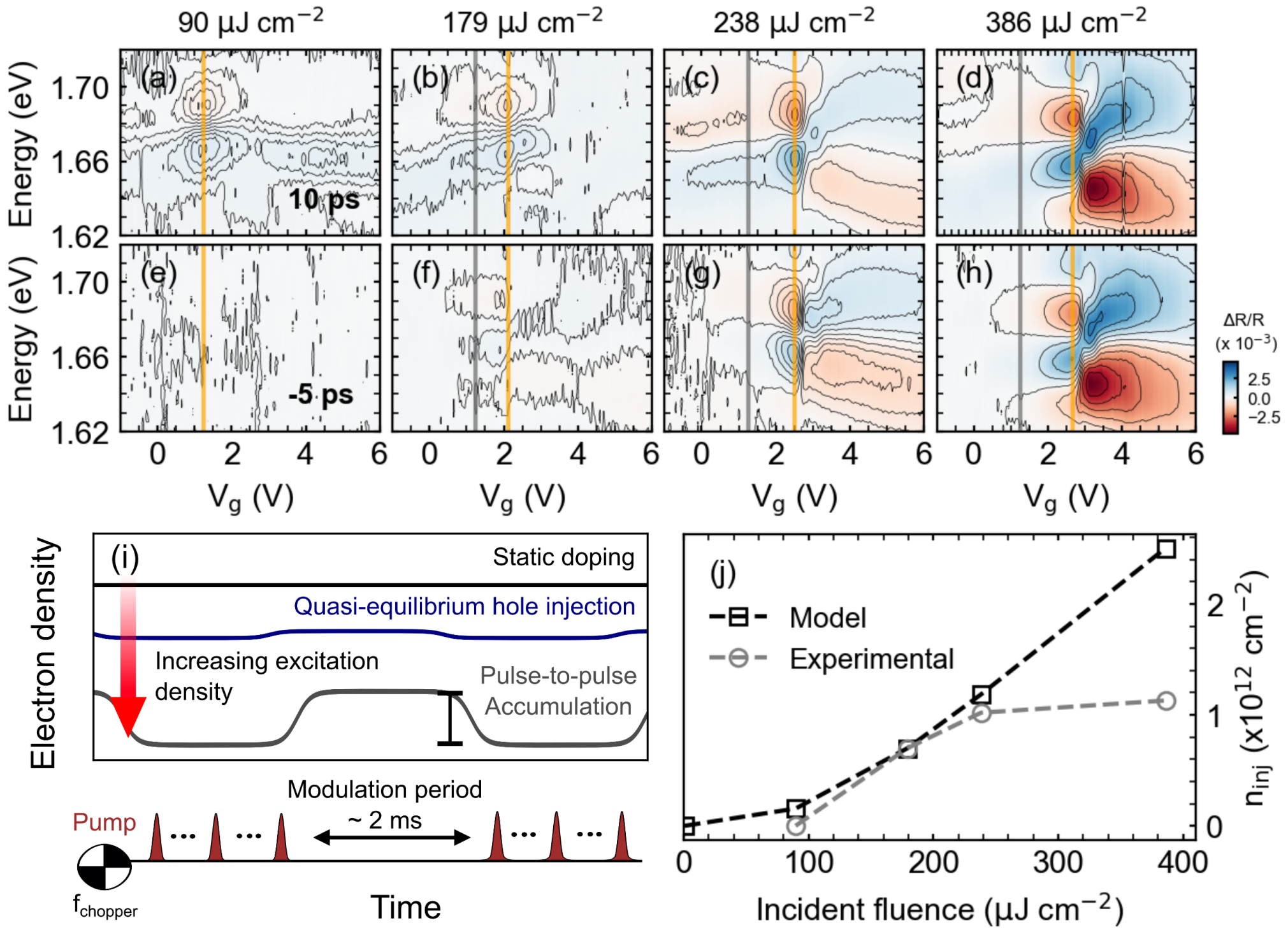

**Figure 2. Spectral signatures of the photo-injected hole population and characteristic timescales.** Fluence-dependent (as labeled for each column) transient reflectance as a function of gate voltage at two different pump-probe delays: Δt = 10 ps in the first column, consisting of panels (a-d), and Δt = -5 ps in panels (e-h). Here, a pump energy of 0.954 eV was employed, and the probe energy was set to cover the lowest energy $tWSe_2$ moiré exciton. The vertical yellow lines in panels (a–h) indicate the positions of the PB signals. The gray lines in panels (b–d) and (f–h) mark the location of the original PB signal in the absence of photo-thermionic injection, shown for comparison. (i) Schematic representation of the doping density at three different excitation fluence conditions. At low fluence, doping is dominated by electrostatic doping due to the applied voltage (black line). As the number of photoinjected holes increases, the total electron doping decreases due to recombination, which shows a quasi-equilibrium hole density condition (blue line). When the fluence is sufficiently high, there is a pulse-to-pulse accumulation resulting in a pump-on/off modulation of the electron density (gray line). (j) Comparison between the experimental injected carriers estimated from the voltage shift of the $\nu_m = 1$ feature trend and the expected hot hole density above the energetic barrier. In subfigures (e) and (f), the contour lines divide the 2D map into three equidistant segments to minimize the visualization of noise. Measurements are performed at T = 4.6K.

The first observation of the systematic shift of the $\nu_m$ = 1 PB feature with ρ reveals that additional static electron doping is required to achieve the same electron filling density. Holes from thermionic injection must be compensated because the injected holes recombine with electrostatically doped electrons. Notably, the persistence of a PB (rather than a sign-flipped PIA) feature at a fixed $V_g$ value indicates that the observed state appears at the same gate voltage during both pump-on and pump-off conditions. If the filling condition were instead transiently shifted between pump-on and pump-off cycles, a sign-flipped PIA spectral feature would emerge. The presence of a long-lived PB signal thus points to the injection of a quasi-steady-state carrier population that persists for the full duration of the pump modulation cycle $\tau_{mod}$ = 2 ms, as illustrated schematically in Fig. 2i. Due to the large contact resistance in the circuit, the discharge of photo-injected hole from Gr, i.e., electrostatic rebalancing, occurs over timescales on the order of $\tau_{res}$ ~10 ms [1], which is significantly slower than the modulation period of the pump beam. From the shift in the $\nu_m$ = 1 PB feature with ρ, we calculate the residual hole-doping density from the photo-thermionic emission process within the $\tau_{mod}$ = 2 ms modulation time window, shown as grey circles as a function of ρ in Fig. 2j. The experimental hole-doping density are in excellent agreement with those from model calculation (black squares, Supporting Information [25]) for ρ = 90- 238 μJ/cm$^2$. Deviation is seen at higher excitation fluence (ρ = 386 μJ/cm$^2$) where the residual hole density from experimental measurement levels off, suggesting accelerated electron-hole recombination in photo-excited fl-Gr at higher ρ.

The second key observation is the emergence of a pre-time-zero PB signal in the transient reflectance with increasing pump fluence, as shown in Figs 2e-h. In pump-probe measurements, pre-time-zero features arise from processes with recovery timescales longer than the laser repetition period $\tau_{rep}$, which is 2.5 μs for the experiments shown in Fig. 2, but slower than the pump modulation period, $\tau_{mod}$. In such cases, the probe pulse effectively senses the residual effect of the preceding pump pulse, creating a signal at negative time delay. To characterize the lifetime of this long-lived response, we conducted repetition rate-dependent measurements, Fig. 3, at a fixed fluence (1.46 eV, 127 μJ cm$^{-2}$) for $\nu_{rep}$ = (a) 400 kHz, (b) 80 kHz, and (c) 50 kHz. The pre-time zero signal decreases with decreasing $\nu_{rep}$ and becomes barely observable for $\nu_{rep}$ = 50 kHz, implying a recovery timescale $\tau_{rec}$ ~ 20 μs. Since this timescale is approximately three-orders-of-magnitude shorter than the timescale $\tau_{res}$ ~10 ms for the discharging of residual photo-injected holes, we attribute $\tau_{rec}$ to a different origin, likely the lateral diffusion of electrons into the region of depleted electron density from recombination with photo-injected holes in the focused laser spot (diameter ~1.4 μm and 2.1 μm at hν = 1.46 eV and 0.954 eV, respectively). This interpretation is supported by the order-of-magnitude estimation. The low temperature hole mobility in bilayer $WSe_2$ is $\mu$~$10^3$ $cm^2V^{-1}s^{-1}$ [32]. We expect the electron mobility to be lower based on the effective mass and a further lowering in the presence of the corrugated moiré potential. We assume an electron mobility in the twisted $WSe_2$ bilayer of $\mu \sim 10^2\ cm^2 \cdot V^{-1} \cdot s^{-1}$. Based on the Einstein relationship $D = \mu \cdot k_B \cdot T/e$, where $D$ is the diffusion constant, $k_B$ is the Boltzmann constant and $e$ is electron charge, we estimate an electron diffusion constant of $D \sim 0.04\ cm^2 \cdot s^{-1}$. This gives an estimated diffusion length in 20 μs of $L \sim 9$ μm, which is of the order of the sample size.

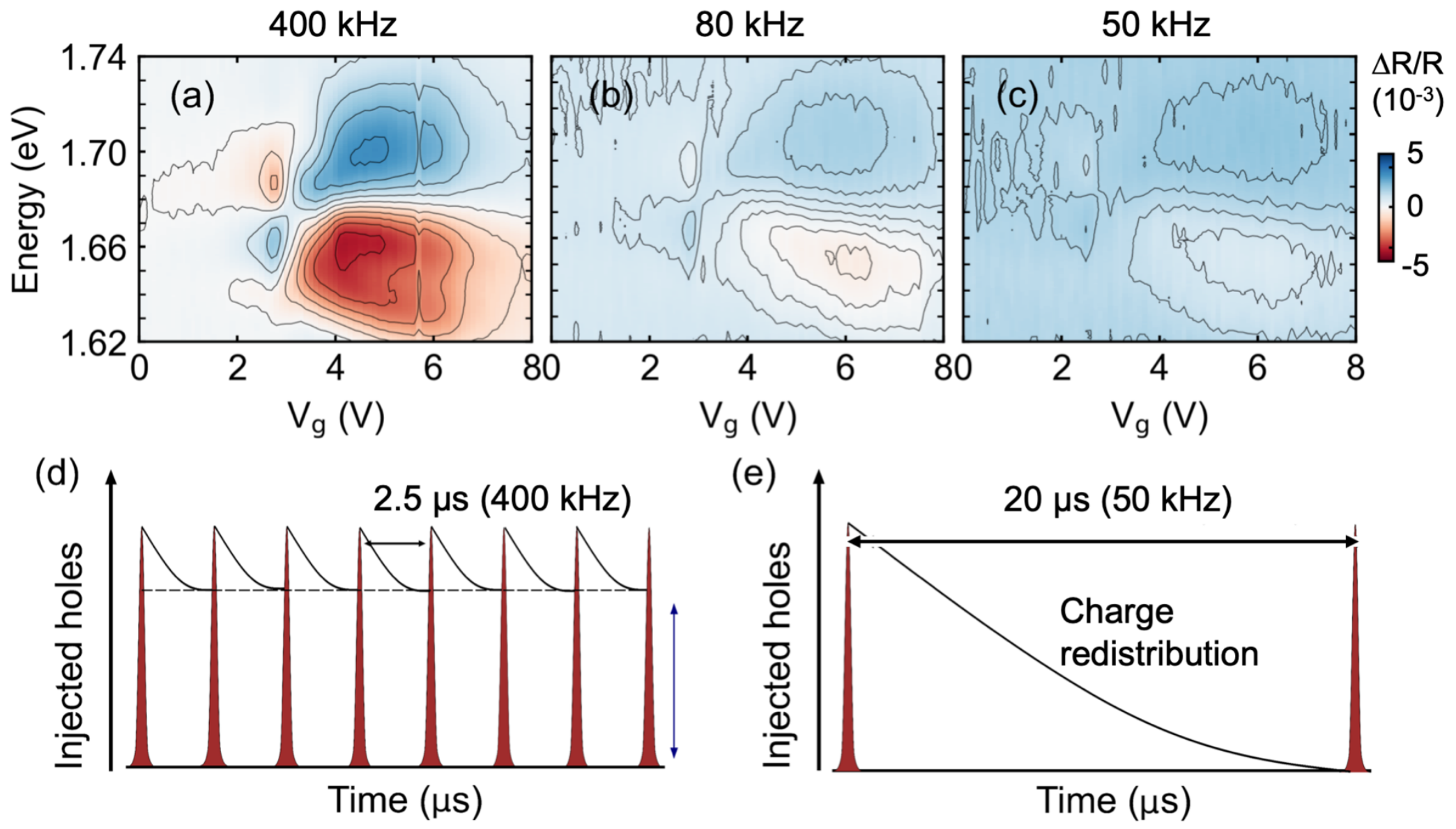

**Figure 3. The repetition rate-dependent transient reflectance response reveals a long-lived local charge imbalance.** Repetition rate-dependent (as labeled for each row) transient reflectance as a function of gate voltage at two different pump-probe delays: Δt = 2.5 μs (-5 ps) in the first column consisting of panels (a), (c), and (e). Here, a pump energy of 1.46 eV (127 μJ $cm^{-2}$) was employed, and the probe energy was set to cover the first $WSe_2$ moiré exciton. Schematic representation of the charge redistribution timescale as related to laser repetition periods of (d) 2.5 μs and (e) 20 μs. Measurements are performed at T = 4.6K.

Finally, we observe the emergence of a sign-flipped feature in the transient reflectance signal at $V_g$ above that corresponding to the $\nu_m = 1$ state, at the high excitation fluences: ρ = 238 and 386 μJ/$cm^2$, (Fig. 2g, 2h). This sign-flipped PIA-like response indicates an increase in the exciton oscillator strength following pump excitation. There is a net reduction in electron doping during the pump-on state, arising from photo-thermionic hole injection from the fl-Gr gates that recombine with electrostatically doped electrons. The moiré system with electrostatically doped electron density in the conducting state at $\nu_m > 1$ can be transiently shifted towards the $\nu_m = 1$ insulating state by photo-thermionic emitted holes from the Gr electrodes, thus, reducing the dielectric constant and giving rise to the observed PIA response. This photo-induced metal-to-insulator transition is established in more detail for the $WS_2/WSe_2$ moiré system in the companion manuscript [26]. In addition, this reduction of the net doped electron density also decreases the dielectric constant and might contribute to the PIA-like response.

**Photonic control of photo-thermionic hole injection**

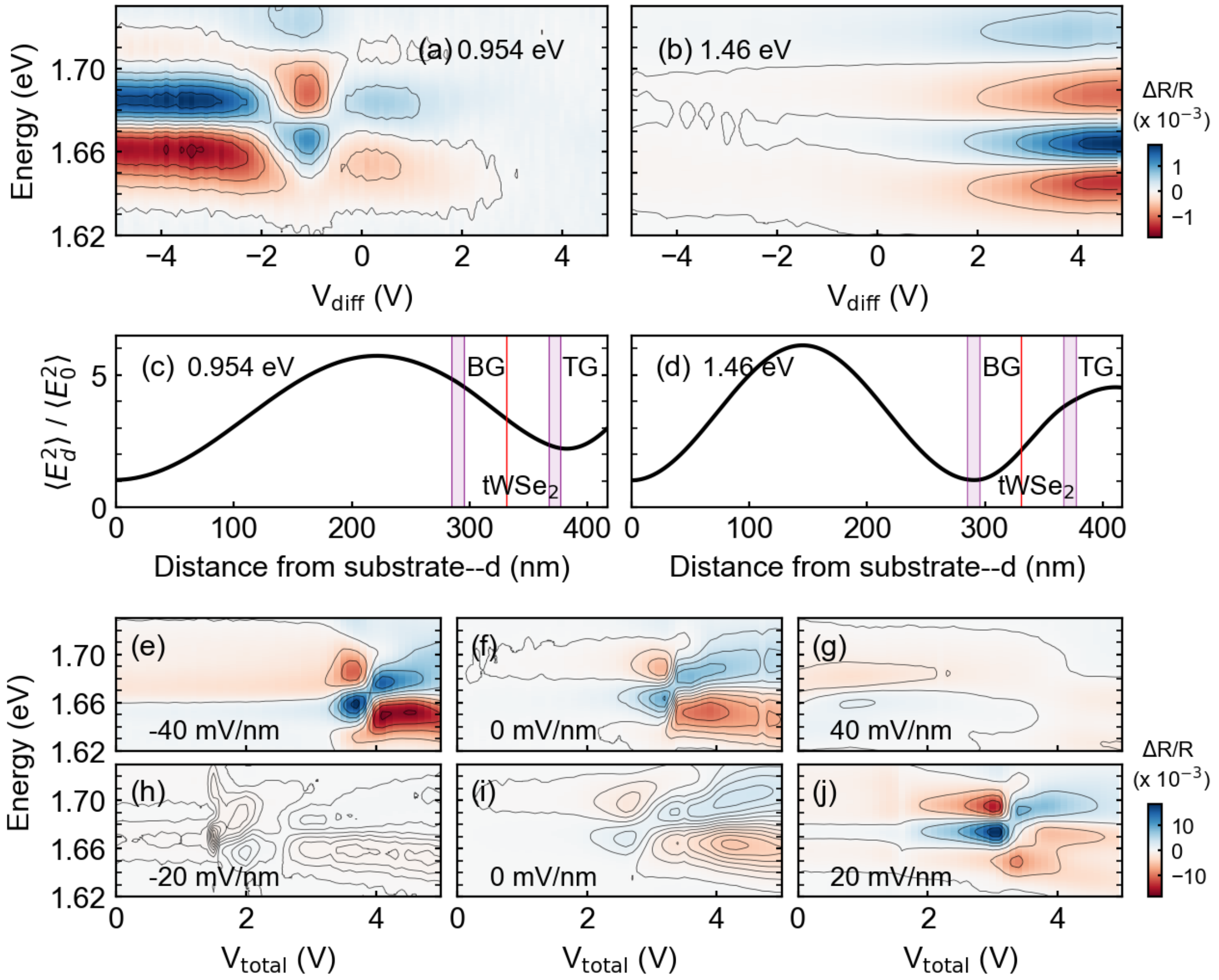

**Figure 4. Tuning photo-thermionic injection with varying pump energy and displacement field.** (a-b) Pump energy-dependent (as labeled for each row) transient reflectance as a function of voltage difference, ( $V_{diff} = V_{tg} - V_{bg}$) at a pump-probe delay of: $\Delta t$ = -5 ps. Pump fluence in panel (a) is 318 $\mu J/cm^2$ and pump fluence in panel (b) is 127 $\mu J/cm^2$. The transfer matrix model simulated the optical mean square electric field propagating through the device [34], (c) and (d) show the field intensity as a function of distance from the substrate (Si) for an excitation wavelength of 0.954 eV and 1.46 eV, respectively. Transient reflectance spectra as a function of total voltage ( $V_{total} = V_{tg} + V_{bg}$) measured at a delay time of 10 ps with an excitation energy of 0.954 eV (e-g) and 1.46 eV (h-j) at distinct electric field conditions as labeled for each panel. Measurements are performed at T = 4.6K.

The dual-gate device structure allows independent control of the electrostatic doping level and the out-of-plane electric field. The net doping level is given by the sum of the top and bottom gate voltages $V_{tot} = (V_{tg} + V_{bg})/2$, while the vertical electric field is determined by their difference $E \propto V_{diff} = (V_{tg} - V_{bg})$. The multilayer photonic structure can give rise to very different pump laser fields at the top and bottom Gr electrodes, as revealed by the asymmetric response of transient reflectance as a function of $V_{diff}$. Consistent with a photonic effect [33,34], the asymmetry reverses as the excitation photon energy is changed from hν = 0.954 eV to 1.46 eV, Figs 4a-b. We perform a transfer matrix simulation of the optical field propagating through the device at the two photon energies (Figs 4c-d). The simulation confirms that the optical field at the bottom Gr electrode is higher than that at the top Gr electrode at hν = 0.954 eV and this relative field distribution reverses for the two Gr electrodes at hν = 1.46 eV. For the 0.954 eV pump, an electric field oriented from the bottom gate toward the bilayer enhances photo-thermionic injection, consistent with increased hole injection under negative out-of-plane bias (Fig. 4a). In contrast, with 1.46 eV excitation, the dominant absorption in the top gate reverses this behavior, favoring hole injection under positive electric field conditions (Fig. 4b).

This photonic effect is further supported by transient reflectance as a function of gate voltage under different electric fields. Fig 4e-g presents gate-dependent transient reflectance measurements under positive, zero, and negative out-of-plane electric fields with 0.954 eV excitation. Under a negative field, enhanced hole injection leads to a stronger signal and a shift in the $\nu_m = 1$ insulating state toward higher doping, consistent with increased competition between electrostatic electron doping and photoinjected holes. In contrast, Figs 4h-j show the same set of electric field-dependent measurements under 1.46 eV excitation, where the opposite trend is observed. In this case, the signal is enhanced under a positive electric field, confirming the gate-selective behavior of photo-thermionic injection as determined by the excitation energy and the substrate. These results demonstrate that both the excitation energy and the applied electric field direction serve as tunable parameters to control thermionic injection and thereby modulate the charge density of moiré systems.

## Conclusion

We demonstrate photodoping of a dual-gated $tWSe_2$ device via thermionic hole injection from the graphite gates, as evidenced by three distinct spectroscopic signatures. The injected hole population can be selectively controlled by the photon energy, incident fluence, and out-of-plane electric fields. Photo-thermionic hole injection occurs on sub-picosecond timescales—orders of magnitude faster than conventional electrostatic gating, which typically operates on millisecond-to-second timescales. Our work provides a detailed and quantitative understanding of photo-thermionic injection and establishes photo-thermionic injection in van der Waals semiconductor heterostructures as an ultrafast and tunable mechanism for controlling carrier density in active layers in vdW structures. Furthermore, this work highlights the potential of utilizing the full vdW device stack to probe and engineer the non-equilibrium dynamics of quantum phases in two-dimensional materials.

## Methods

### Device fabrication

A $WSe_2$ monolayer was mechanically exfoliated from flux-grown bulk crystals. The monolayer was stacked as a bilayer by the dry transfer "tear and stack" technique with a polycarbonate stamp. The sample is grounded via Gr contacts connected to the homobilayer and single-crystal hBN dielectrics, and graphite gates are used to encapsulate the device and provide control of the carrier density and displacement field (if desired) via external source meters (Keithley 2400). The moiré period are confirmed through piezo force microscope (PFM). All electrodes are defined with electron beam lithography and made of a three-layer metal film of Cr/Pd/Au (3 nm/17 nm/60 nm).

### Steady-state reflectance

All optical measurements were performed in a vacuum environment ($<10^{-7}$ torr) using a closed-cycle helium cryostat (Fusion X-Plane, Montana Instruments) to maintain the sample at cryogenic temperatures. For steady-state reflectance, broadband illumination was provided by a 3200 K halogen lamp (KLS EKE/AL). The light was collimated and focused onto the sample using a 100× objective lens with a numerical aperture of 0.75. Reflected light from the sample was collected through the same objective, directed into a grating spectrometer, and detected with a CCD camera (Blaze, Princeton Instruments). The reflectance spectrum was obtained by normalizing the reflected intensity from the sample to that of a reference region.

Time-resolved differential reflectance spectroscopy

Ultrafast transient reflectance measurements were performed using a femtosecond laser system (Pharos, Light Conversion) operating at 1030 nm with an 89 fs pulse duration, 400 kHz repetition rate, 10 W average power, and 20 μJ pulse energy. The laser output was split into pump and probe beams. To generate the probe, a portion of the fundamental beam was focused into a YAG (Yttrium Aluminum Garnet) crystal to produce a white-light continuum. This continuum was spectrally filtered using a 700 nm long-pass and 800 nm short-pass filter to isolate the energy window corresponding to the lowest $WSe_2$ moiré exciton. The pump beam was sent through a tunable optical parametric amplifier (Orpheus-NEO, Light Conversion, 315–2700 nm) to produce the desired excitation wavelength. It then passed through a motorized delay stage to control the pump-probe time delay (Δt) and an optical chopper for modulation between pump-on and pump-off conditions. Pump and probe beams were recombined and focused onto the sample through a 100×, 0.75 NA objective lens. The pump and probe spot sizes at the sample were ~2.1 μm and ~1.38 μm in diameter, respectively. Reflected probe light was collected by the same objective, filtered to remove residual pump light, dispersed by a spectrometer, and detected using a CCD array (Blaze, Princeton Instruments). The transient reflectance signal (ΔR/R) was calculated from the difference between the pump-on and pump-off spectra at various delay times.

Assignment of the filling factors

The charge density, in the $tWSe_2$ heterostructure, controlled by the applied gate voltages ($V_t$ and $V_b$ for the top and bottom gates, respectively), is determined using the parallel-plate capacitor model: $n = \frac{\varepsilon\varepsilon_0 \Delta V_t}{d_t} + \frac{\varepsilon\varepsilon_0 \Delta V_b}{d_b}$, where $\varepsilon \approx 3$ is the out-of-plane dielectric constant of hBN, $\varepsilon_0$ is the permittivity of free space, $\Delta V_i$ is the applied gate voltage, and $d_i$ is the thickness of the hBN spacer. The moiré density, $n_0$, is determined from the moiré lattice constant, $a_M$ according to $n_0 = \frac{2}{\sqrt{3}a_M^2}$. The filling factor, $\nu = \frac{n}{n_0}$, is fit based on the experimental gate-dependent steady state and transient reflectance measurements. Where applicable (D1 and D2), the calculated filling factors are compared to those obtained based on the $a_M$ measured from PFM ($a_M \approx 10.3$ nm for D1 and $a_M \approx 10.6$ nm for D2). For D1, the thickness of the top and bottom hBN spacers is determined to be $d_t \approx 41.3$ and $d_b \approx 40.3$ nm, respectively. The twist angle is determined to be ~58.1$\pm$0.1°. Similarly, for device, D2, the thickness of the top and bottom hBN spacers is determined to be $d_t \approx 47.8$ and $d_b \approx 46.3$ nm, respectively, and the twist angle is determined to be ~58.2$\pm$0.1°. The twist-angle uncertainty is extracted from the filling-factor calibration uncertainty in static reflectance gate scan for the $tWSe_2$ bilayer.

## Acknowledgments

This work was supported by the US Department of Energy Office of Basic Energy Sciences (DOE-BES) under award DE-SC0024343. Device fabrication and characterization were supported by the NSF Materials Research Science and Engineering Center under award DMR-2011738. Development of the photo-thermionic emission mechanism was supported by Department of Defense Multidisciplinary University Research Initiative (MURI) grant number W911NF2410292. E.R-G. acknowledges support from the Columbia Quantum Initiative Postdoctoral Fellowship. E.A.A. gratefully acknowledges support from the Simons Foundation as a Junior Fellow in the Simons Society of Fellows (965526).

## Author contributions

Y.L. prepared the twisted bilayer devices. Y.L., E.R-G., and E.A.A. carried out the optical measurements and analysis. L.H. prepared the TMD single crystals under the supervision of J.H. Y.G., B.Y., and D.S. assisted with the preparation of the devices under C.D.'s supervision. X.Y.Z. supervised the project. The first draft was prepared by Y.L, E.R-G., E.A.A., and X.Y.Z. in consultation with all other authors. All authors read and commented on the manuscript.

## Data availability statement

All data supporting the conclusions of this manuscript are available from the authors upon reasonable request.

**SM: Ultrafast Charge-Doping via Photo-Thermionic Injection in van der Waals Devices**

Yiliu Li, Esteban Rojas-Gatjens, Yinjie Guo, Birui Yang, Dihao Sun, Luke Holtzman, Juseung Oh, Katayun Barmak, Cory R. Dean, James C. Hone, Nathaniel Gabor, Eric A. Arsenault, Xiaoyang Zhu

**Supplementary notes:**

<u>Simple model for photodoping temperature dependence on power and voltage.</u>

Ma et al. [24] describe the photocurrent of a Graphene/hBN/Gr device by considering contributions from tunneling and above-barrier thermionic emission. The interlayer transmission probability for carriers with an energy above the barrier is 1, while the tunneling transmission probability is $\propto e^{-d}$ where d is the thickness of the barrier (hBN). We discard tunneling as a photodoping mechanism since we used thick hBN as spacers (40 nm). The number of holes with energies above the energy barrier ($\Delta_b$) described by equation (1). Where $DoS(E)$ is the density of states of the few-layer graphene gate, $f_{FD}(E, T_e, E_F)$ is the Fermi-Dirac distribution, $T_e$ is the electronic temperature, and $E_F$ is the Fermi energy. Note that control over the Fermi energy is effectively a control over the energy barrier ($\Delta_b$).

$$n_{therm} = \int_{\Delta_b}^{\infty} dE\ DoS(E) \cdot f_{FD}(E, T_e, E_F) \quad (1)$$

We take: $DoS(E) = \frac{2|E|}{\pi(\hbar v_f)^2}$ and $f_{FD}(E) = \frac{1}{e^{(E-E_F)/k_bT}+1}$. Then the integral in (1) yields.

$$n_{h+} = \frac{2}{\pi(\hbar v_f)^2} \int_{\Delta_b}^{\infty} dE\ E \cdot \left(\frac{1}{e^{(E-E_F)/k_bT}+1}\right) \quad (2)$$

For the condition of $(\Delta_B - E_F) \gg k_bT$, the expression simplifies to [19]:

$$n_{h+} \approx \frac{2}{\pi}\left(\frac{k_bT}{\hbar v_F}\right)^2 \left[\frac{\Delta_B}{k_bT} + 1\right] e^{-(\Delta_B - E_F)/k_bT} \quad (4)$$

Similarly, $\frac{\Delta_B}{k_bT} \gg 1$, therefore:

$$n_{h+} \approx \frac{2\Delta_B k_b T_e}{\pi(\hbar v_F)^2} e^{-(\Delta_B - E_F)/k_bT_e} \quad (4)$$

We connect this expression to the experimental variables of laser fluence and voltage. From the fluence, we can estimate the electronic temperature following:

$$2\int_0^\infty dE\ DoS(E)\cdot f_{FD}(E,T_e,E_F)\cdot E = F_{photon} \qquad (5)$$

Assuming graphene's density of states:

$$T = \left(\frac{\pi(\hbar v_F)^2 F_{photon}}{(4\cdot 1.803\cdot k_b)^3}\right)^{1/3}$$

$$E_F = \hbar v_F\sqrt{\pi}\cdot\sqrt{n_{doping}} \quad \text{and} \quad n_{doping} = \frac{\varepsilon\varepsilon_0\Delta V_t}{d_t} \qquad (6)$$

We use these simple equations to map the injected carriers in the fluence and voltage space, presented in figure 1 and 2 of the main text. We use only one calibration parameter, corresponding to the $\beta$ such that $F_{photon} = \beta F_{incident}$. The rest of the parameters are taken from the literature under the assumption that a few-layer graphite will behave similarly to graphene.

Transfer matrix model

The electric field resulting from the propagation of a coherent electromagnetic plane wave through the "thin-film" layered structure was simulated following the conventional Transfer-matrix method in matrix form as described by Hansen [34]. Effectively, the device is represented as a five-layer structure (hBN/fl-Gr/hBN/fl-Gr/$SiO_2$) where the Si is assumed as an infinite output medium. Let us summarize the main equations for an N-layer general case, specifically for TE polarization.

The input and output electric fields are related by equation (1).

$$Q_1 = \begin{bmatrix} E_{in}/E_{out} \\ H_{in}/E_{out} \end{bmatrix} = \begin{bmatrix} U_1 \\ V_1 \end{bmatrix} = \prod_i M_i \begin{bmatrix} 1 \\ n_s \end{bmatrix}$$

Where the leftmost matrix corresponds to the first film that light traverses. $M_i$ corresponds to the characteristic matrix defined as equation (2), with $\delta = 2\pi nd/l$ where $n$ is the refractive index.

$$M_i = \begin{bmatrix} \cos(\delta) & i\cos(\delta)/n \\ in\sin(\delta) & \cos(\delta) \end{bmatrix}$$

To calculate the fields, we define the inverse matrix of the characteristic matrix, $NM = I$. With which we can extract the field at a general point as:

$$\begin{bmatrix} U_k(z) \\ V_k(z) \end{bmatrix} = Q_k(z) = N_k(z) \prod_{j=i-k} N_j \begin{bmatrix} U_1 \\ V_1 \end{bmatrix}$$

Then, assuming the $|E_{out}| = 1$, the mean-square electric field at a general point can be calculated relative to the output mean-square electric field as:

$$\langle E(z)^2 \rangle \langle E(0)^2 \rangle = \frac{1}{4} |U_k(z)|^2$$

**Supplementary figures:**

<u>Reproducibility of main results across two devices.</u>

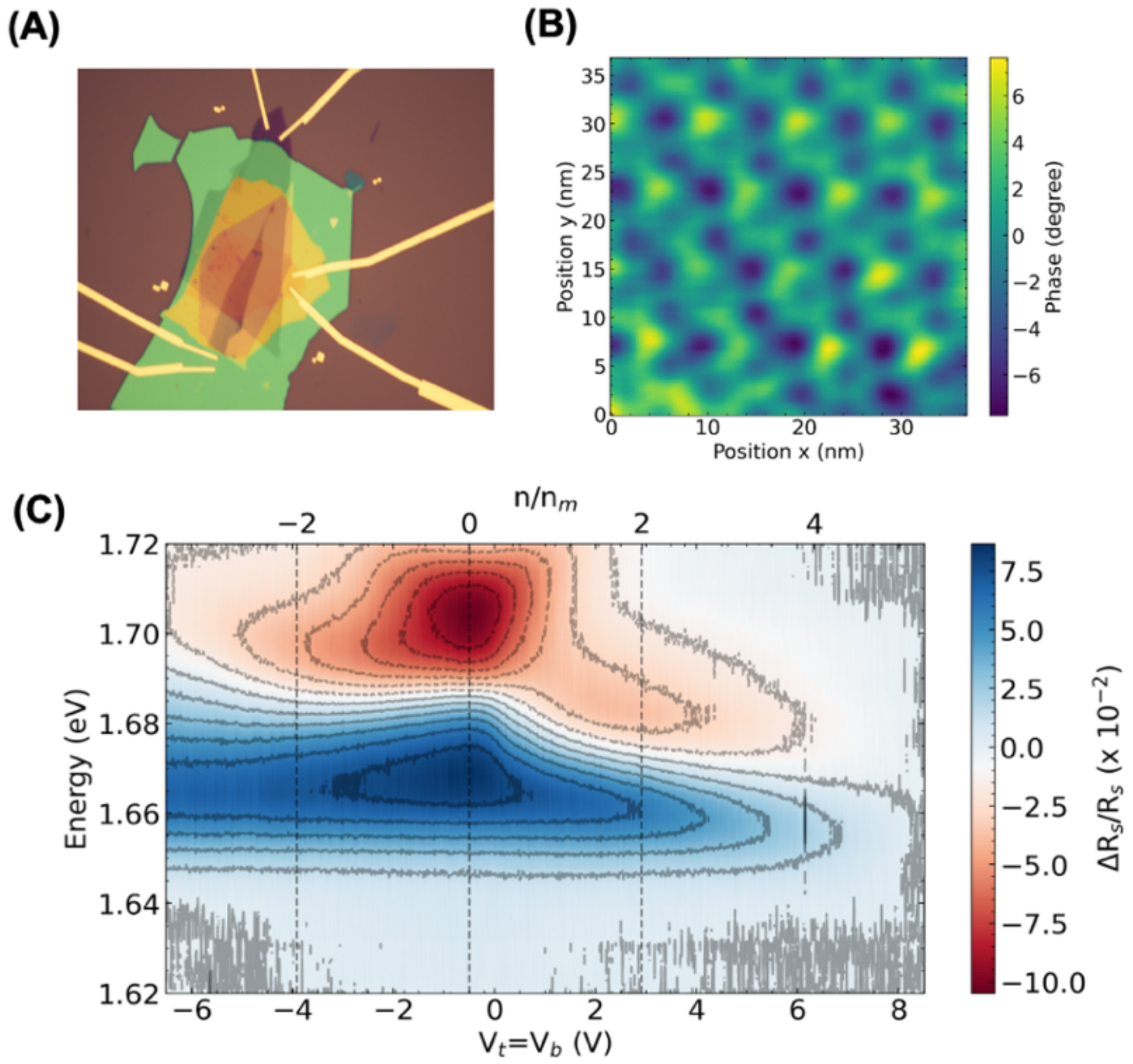

**Figure S1.** Device image and characterization for device D2. (A) microscope image of D2. (B) piezo force microscope image of moiré landscape of D2. (C) Static white light reflectance map as a function of gate voltage.

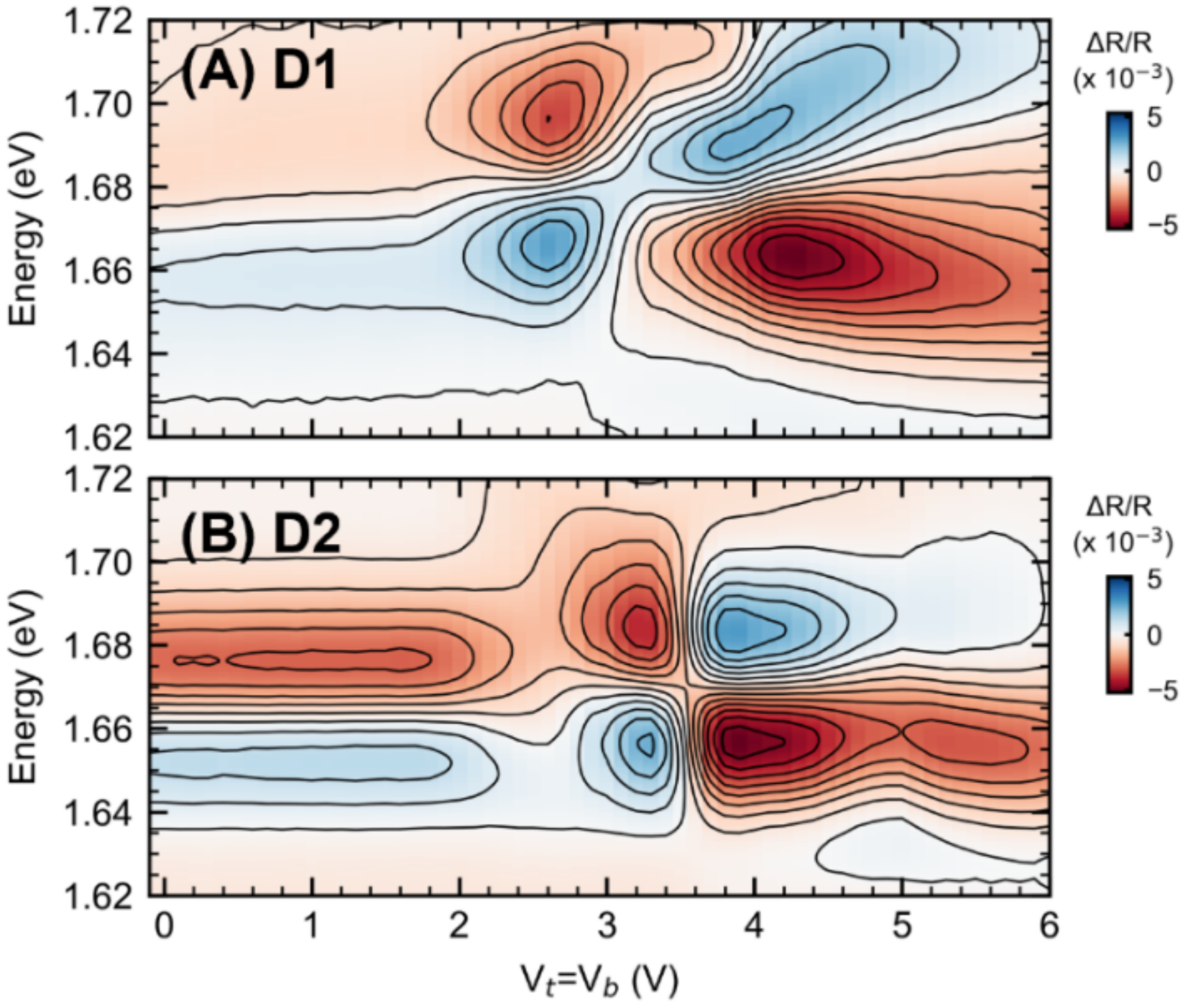

**Figure S2.** Transient reflection reflectance gate maps response shown as a function of doping fluence (as labeled for each row) at a pump-probe waiting times: Δt = 10 ps for two distinct devices with similar twist angles, (A) D1 and (B) D2. Here, a pump energy of 1.46 eV was employed and the probe energy was set to cover the lowest energy, and we are probing the first $WSe_2$ moiré exciton.